# Correlation-driven quantum geometry effects in a Kondo system

Ruizi Liu[1*], Zehan Chen[2*], Xingkai Cheng[2], Xiaolin Ren[2], Yiyang Zhang[2], Xuezhao Wu[1], Chengping Zhang[2], Kun Qian[1], Ching Ho Chan[1], Junwei Liu[2,3], Kam Tuen Law[2], Qiming Shao[1,2,3†].

[1]Department of Electronic and Computer Engineering, The Hong Kong University of Science and Technology, Clear Water Bay, Kowloon, Hong Kong SAR 999077, China

[2]Department of Physics, The Hong Kong University of Science and Technology, Clear Water Bay, Kowloon, Hong Kong SAR 999077, China

[3]IAS Center for Quantum Matter, The Hong Kong University of Science and Technology, Hong Kong, China

*Equal contribution

†Email: eeqshao@ust.hk

Quantum geometry, including quantum metric and Berry curvature, which describes the topology of electronic states, can induce fascinating physical properties [1,2]. Symmetry-dependent nonlinear transport has emerged as a sensitive probe of these quantum geometric properties [3–7]. However, its interplay with strong electronic correlations has rarely been explored in bulk materials, particularly in a Kondo lattice system. Here, we uncover correlation-driven quantum geometry in centrosymmetric antiferromagnetic iron telluride (FeTe). We experimentally observe the quantum metric quadrupole-induced third-order nonlinear transport, whose angular dependence reflects magnetic structure in FeTe. The nonlinear transport signals follow Kondo lattice crossover and vanish at high temperatures. Our theory suggests that a Kondo lattice formed at low temperatures explains the emergence of quantum geometry, which is induced by the opening of a hybridization gap near the Fermi energy. This discovery establishes a paradigm where quantum geometry arises not from static symmetry breaking but from dynamic many-body effects and provides a zero-field probe for sensing antiferromagnetic order.

## Introduction

Quantum metric and Berry curvature encode the quantum geometry information of the electronic bands of materials and they can induce novel physical phenomena [1], [2]. In recent years, breakthroughs have been made in the study of the quantum geometry of itinerant electrons through nonlinear transport measurements [3–5,8,9]. Unlike the intrinsic anomalous Hall effect which depends on the Berry curvature monopole [10,11] (which is the sum of the Berry curvature of the occupied states), nonlinear transport can capture the properties of the higher order moments of the quantum geometry such as the dipoles and quadrupoles of Berry curvature and quantum metric [3,5–7,12–15]. Generally, the quantum geometric moments and associated physical manifestations are closely related to symmetries. For example, to achieve finite Berry curvature dipole (BCD) or quantum metric dipole (QMD) and a measurable second-order nonlinear transport, the inversion symmetry ($P$) has to be broken by the low crystal symmetry [3,16], strain [17,18], magnetic order [4,5] or heterostructures [19]. Similarly, Berry curvature quadrupole (BCQ), which can contribute to the third-order nonlinear Hall response, is finite only when the time-reversal symmetry ($T$) is broken and in materials with specific magnetic point group symmetry [7,12,20]. Recently, quantum metric quadrupole (QMQ) induced third-order nonlinear effects were observed in antiferromagnets, which was sensitive to the applied magnetic field and the spin order of the material [12]. Therefore, the nonlinear response offers a powerful way to probe quantum geometry and symmetry properties of materials (Fig. 1a).

After intense experimental investigations in recent years, non-interacting systems have already exhibited a well-established interplay between quantum geometry and nonlinear transport [3–5,7]. To generate sizable nonlinear responses by quantum geometry, previous studies utilized materials with nontrivial topological bands such as semimetals [16,21,22] and topological insulators [4,5,12]. Another efficient method to modulate quantum geometry is utilizing correlation effects, which have also been studied in a few thin film systems, such as in monolayer graphene [23], twisted bilayer graphene (t-graphene) [17], [18], and twisted bilayer $WSe_2$ (t-$WSe_2$) [24]. Recently, a second-order nonlinear Hall response was observed in a nonmagnetic Weyl-Kondo semimetal [25] and a second-order longitudinal nonlinear resistance was reported in the strongly correlated oxide [26]. In these correlated systems, the emergence of second-order nonlinear transport requires inversion symmetry breaking and finite Berry curvature dipole or quantum metric dipole. Therefore, whether electronic correlations can generate higher-order quantum-metric moments and third-order nonlinear transport in centrosymmetric materials, remains unexplored.

In this work, we report the third-order nonlinear response in centrosymmetric antiferromagnet tetragonal FeTe. First, we observed third-order nonlinear longitudinal and transverse responses when an AC current is applied. Importantly, the third order is the leading order response as both the first-order and second-order Hall responses vanish due to $\mathcal{P}$ and $T \circ \tau_{1/2}$ ($\tau_{1/2}$ is the half lattice translation along the $a$-axis). Second, a two-fold rotational symmetry of the nonlinear responses with respect to the crystal axis is observed. This is consistent with the symmetry of the magnetic order of FeTe in the Kondo lattice phase. Third, the observed nonlinear responses persist up to around 80 K where the partial Kondo screening vanish. This indicates that the nonlinear response is related to the Kondo lattice phase. Finally, by an effective Kondo lattice Hamiltonian, we calculated the nonlinear responses induced by the quantum metric quadrupole of the flat bands. The flat band is generated by the hybridization of the itinerant electrons and the local magnetic moments. The effective Kondo lattice model captures both the symmetry properties and the order of magnitude of the nonlinear responses. Therefore, we suggest that the Kondo interaction between itinerant electrons and local moments provides a possible microscopic explanation for the observed nonlinear responses in FeTe. In this work, we highlight the manipulation of the quantum geometrical effects by Kondo physics, which can be probed through nonlinear transport measurements.

## Crystal symmetry and Kondo correlation in FeTe

FeTe is widely studied as the parent compound of iron chalcogenide superconductors [27] and represents a prototypical Hund metal [28,29]. As shown in Fig. 1b, the crystal structure transitions from $P4/nmm$ symmetry to $P2_1/m$ symmetry below Néel temperature ( $T_N$~60 K [30]), accompanied by the formation of bicollinear antiferromagnetic order [31], [32]. In the spin-ordered phase, one mirror line is preserved along the $a$-axis while four mirror lines are present in the non-magnetic phase [33] (Fig. 1b). Crucially, although under $P$ and $T(\circ\ \tau_{1/2})$ symmetry, the QMQ-induced third-order nonlinear transport remains allowed across all structural phases [34]. Meanwhile, Kondo lattice in FeTe at low temperatures (Figs. 1c-d) was observed in previous reports, featuring a partial screening of local magnetic moments by itinerant electrons and the formation of hybridization gap formation as confirmed by neutron scattering [35] and angle-resolved photoemission spectroscopy (ARPES) [36], respectively. These strong correlation effects strongly modify the band structure of the material, which is expected to affect the quantum geometric effects as we demonstrate below. Detailed descriptions of magnetic order and Kondo physics in FeTe are presented in Supplementary Note 1.

## Observation of the third-order nonlinear transport in FeTe

To study symmetry- and Kondo physics-dependent transport properties of FeTe, we fabricated devices (see Supplementary Notes 2-4 for characterization) into a disk shape integrated with 8 electrodes (Fig. 2a). First, we conduct longitudinal resistance measurements (Fig. 1d). At low temperatures, we observe a Fermi liquid characteristic as $\frac{d\rho}{dT} \propto T$ below ~20 K. An abrupt drop of resistance occurs ~57 K, which is associated with the formation of antiferromagnetic order at Néel temperature $T_N$. These transport features are consistent with previous reports [36], [37]. Subsequently, at 10 K, under an AC current with a frequency of 19.357 Hz, we detected dominant third-harmonic longitudinal and transverse voltages with negligible second-harmonic signals (Fig. S7). The second harmonic signals are absent in the presence of $\mathcal{P}$ [34]. Linear anomalous Hall effect vanishes as well (Fig. S8d) due to $\mathcal{T} \circ \tau_{1/2}$.

Then we carried out angle-dependent nonlinear transport measurement at 20 K (monoclinic phase) by applying AC current along different crystal directions. A current $I$ was injected into the disk-shaped device at angle $\theta$, which is defined as the angle between the current and the $x$-axis (Figs. 2a-b). Note that the $x$-axis corresponds to the $a$-axis, with the misalignment between electrode and crystal direction calibrated via resistance anisotropy, crystalline planar Hall effect, and anisotropic magnetoresistance measurements (see Supplementary Notes 3 and 4). As shown in Figs. 2c-d, third-order nonlinear voltages $V_{\parallel}^{(3)}$ and $V_{\perp}^{(3)}$ for different $\theta$ at 20 K are plotted against the cubic of the first-harmonic longitudinal voltage $V_{\parallel}$. The cubic relationship between nonlinear voltages and the first harmonic longitudinal voltage $V_{\parallel}$ is observed at all angles, confirming the third-harmonic characteristic. Importantly, the two-fold dependence of $\theta$ in the third-order signal (Fig. 2e) is consistent with $P2_1/m$ symmetry of the sample. When current is applied along or perpendicular to mirror line of FeTe crystal, the third-order nonlinear transverse signals disappear while the longitudinal counterpart approaches maximum or minimum [34]. To eliminate the effects caused by the shape of the device, we define sample-size-independent quantities: $E_{\parallel}^{(3)} = \frac{V_{\parallel}^{(3)}}{L_1}$, $E_{\perp}^{(3)} = \frac{V_{\perp}^{(3)}}{L_2}$ for the third-order longitudinal and transverse electric fields, respectively, and $E_{\parallel} = \frac{V_{\parallel}}{L_1}$ for the first-order longitudinal electric field, where $L_{1(2)}$ is the distance between the related electrodes. The angular dependence of the nonlinear responses can be captured and using the equations below (see detailed derivations in Supplementary Note 5):

$$\frac{E_{\parallel}^{(3)}}{E_{\parallel}^{3}} = \rho_a \frac{r^3\sigma_{11}\cos^4\theta + 3r(\sigma_{12}+\sigma_{21})\cos^2\theta\sin^2\theta + r^{-1}\sigma_{22}\sin^4\theta}{(r\cos^2\theta+\sin^2\theta)^3}, \tag{1}$$

$$\frac{E_{\perp}^{(3)}}{E_{\parallel}^{3}} = \rho_a \frac{\cos\theta\sin\theta\,[(3r\sigma_{21} - r^3\sigma_{11})\cos^2\theta + (r^{-1}\sigma_{22} - 3r\sigma_{12})\sin^2\theta]}{(r\cos^2\theta+\sin^2\theta)^3}, \tag{2}$$

where $\sigma_{11}$, $\sigma_{12}$, $\sigma_{21}$, and $\sigma_{22}$ are the third-order nonlinear conductivity tensor components allowed materials with $P2_1/m$ symmetry [38], and $r$ is the ratio of resistivity along different crystal directions ($\rho_a/\rho_b$). Excellent agreement between model and data (Fig. 2e) confirms symmetry-controlled nonlinear responses in FeTe.

## Temperature dependence of nonlinear transport and quantum metric mechanism

To elucidate the origin of nonlinear transport in FeTe, we systematically investigate its temperature dependence. We show the third-order nonlinear voltages as a function of temperature in Fig. 3a with a fixed AC current in amplitude of 1 mA. Below $T_N$, both longitudinal and transverse responses gradually decrease with increasing temperature. Then, an abnormal peak emerges near $T_N$, primarily arising from extrinsic origins such as Joule heating effect (see Supplementary Note 6). Following the monoclinic to tetragonal phase transition in FeTe, the third-order transverse response exhibits a sign change resulted from the major carrier-type change [36] (see ordinary Hall results in Supplementary Note 2). Above $T_N$ where the space group of FeTe belongs to $P4/nmm$, we observe two-fold angular dependence in the third-order transport (Fig. S14c), which originates from incommensurate magnetic order-induced symmetry reduction (see Supplementary Note 7). When the temperature exceeds $T^*{\sim}80\ K$ the third-order nonlinear transverse signal disappears. In contrast, the longitudinal signal here does not show consistent angular dependence with crystalline symmetry (Fig. S14d). Moreover, longitudinal signals remain nonzero above $T^*$, which suggests that longitudinal nonlinear signals above $T^*$ may contain a significant Joule heating effect rather than an intrinsic effect (see Supplementary Note 7). Therefore, we conclude that intrinsic nonlinear transport vanishes at high temperatures. Here $T^*$ denotes the characteristic temperature scale below which the correlated Kondo-hybridized state relevant to the nonlinear response becomes observable. This temperature scale is identified experimentally from the disappearance of the nonlinear signal and is further supported by the deviation of the resistivity from its high-temperature$-\ln T$ behavior.

We then apply the scaling law analysis to validate our claim that QMQ contributes to the third-order nonlinear transport in FeTe (see theory details in Supplementary Note 8). The conductivity is modulated through varying temperatures (Fig. 3b). Here, we select the response in the temperature range at low temperatures far from $T_N$, sufficiently far from $T_N$, where the carrier density (see Supplementary Note 2) and the Fermi surface remain approximately unchanged [36]. Using sample-size-independent quantities $\frac{E_{\parallel}^{(3)}}{E_{\parallel}^{3}}$ and $\frac{E_{\perp}^{(3)}}{E_{\parallel}^{3}}$ , we plot these quantities against the square of conductivity, $\sigma^2$ in Figs. 3c-d. The relationships are expected to be [6]:

$$\frac{E_{\parallel}^{(3)}}{E_{\parallel}^{3}} = \xi_{\parallel}\sigma^2 + \eta_{\parallel}, \quad (3)$$

$$\frac{E_{\perp}^{(3)}}{E_{\parallel}^{3}} = \xi_{\perp}\sigma^2 + \eta_{\perp}, \quad (4)$$

where $\xi_{\parallel(\perp)}$ and $\eta_{\parallel(\perp)}$ are Drude and QMQ contributions respectively. Given $\frac{E_{\parallel(\perp)}^{(3)}}{E_{\parallel}^{3}} \sim \frac{\sigma_{\parallel(\perp)}^{(3)}}{\sigma}$ with $\sigma$ proportional to relaxation time $\tau$, Eqs. (3)(4) reveal two distinct contributions to $\sigma^{(3)}$: a $\tau^3$ term (Drude contribution) and a $\tau^1$ term (QMQ contribution [6,34]). We note that this scaling analysis relies on the assumption that the electronic structure and the relevant quantum-geometric quantities are approximately unchanged within the fitting temperature window, so that the dominant temperature dependence can be attributed to $\tau$. This condition is expected to be better satisfied in the selected low-temperature regime, where the drastic Fermi-surface reconstruction near $T_N$ is avoided. Nevertheless, FeTe remains a correlated multiband system even below 30 K. In particular, the $d_{xy}$ orbital-derived coherent spectral weight can still evolve with temperature, which may lead to weak deviations from an ideal $\tau$-controlled linear scaling. We therefore use the scaling analysis as a phenomenological consistency check for separating these two contributions. In Figs. 3c-d, the linear fittings show finite intercepts for both longitudinal and transverse third-order nonlinear response, which proves the QMQ contribution. Although the limited conductivity range in FeTe introduces additional uncertainty into the quantitative extraction of the intercepts, the observed scaling behavior provides important supporting evidence for the QMQ scenario. Combined with the exclusion of major alternative extrinsic origins in Supplementary Note 9, these results support the interpretation that the observed third-order nonlinear transport contains a QMQ contribution. Meanwhile, we compare the nonlinear conductivity from QMQ and Drude contribution in Fig. S33 and characterize the third-order nonlinear response in FeTe as weakly Drude contribution-dominant.

At elevated temperatures, diminished nonlinear transport suggests reduced QMQ in FeTe. Our measurements on other devices yield consistent results (see Supplementary Note 10). Above ~ 80 K, QMQ vanishes alongside negligible nonlinear signal.

## Quantum metric quadrupole induced by Kondo hybridization

Recent experiments have proven the Kondo lattice physics of FeTe at temperature below $T_N$ [35]' [36]. Motivated by this correlation crossover and the strong reliance of third order nonlinear transport on magnetic order, we consider a phenomenological Kondo-hybridization model as a possible microscopic route explaining the observed QMQ response (Fig. 4a). At high temperatures, strongly localized Fe $d_{xy}$ orbital is incoherent and does not interact with other bands [35]' [39]. Thus, the band crossing is trivial and no QMQ presents (Fig. 4b). At low temperatures, Fe $d_{xy}$ orbital exhibits Hund's coherent behavior [29]. As Hund's coupling favors spins of electrons in different orbitals, local moments tends to form [32]' [40]. Then interaction between localized moments and itinerant electrons can induce a hybridization energy gap [36] (Fig. 4c). Previous spectroscopic studies have provided evidence for Kondo hybridization and hybridization-gap formation in FeTe, mainly around the Γ point [36]. Meanwhile, ARPES studies also reported strong quasiparticle renormalization and peak-dip-hump-like features near the M point, indicating strong correlations in this momentum region [41,42]. Motivated by these observations, we consider both the Γ- and M-point Fermi pockets in a Kondo-hybridization framework. Based on previous ARPES results and our orbital-resolved DFT calculations (see Supplementary Note 11), we construct a minimal phenomenological model in which Fe-$d_{xy}$- and Te-$p_z$-derived low-energy states hybridize near the Γ and M points. We note, however, that the available ARPES and DFT results do not provide conclusive evidence for a specific $d_{xy}$-$p_z$ hybridization gap at the M point. We therefore present the M-point hybridization in Figs. 4b and 4c as a schematic illustration of this phenomenological scenario. With hybridization-induced energy gap and QMQ, the third-order nonlinear transport appears.

Kondo lattice physics is also consistent with the angular dependence of nonlinear transport in FeTe with magnetic order. Under the partial Kondo screening scenario, each Te atom gets spin-polarized [43], and couples antiferromagnetically with the neighboring three Fe atoms and ferromagnetically with the remaining Fe atom in FeTe (Fig. 4a), which naturally induces two-fold anisotropy. In the tetragonal phase (above the magnetostructural transition point $T_N$ ), persistent two-fold nonlinear transverse signals (Fig. S14ac) stem from anisotropic magnetic fluctuation [40,44,45]. Thus, our results suggest that symmetry lowering in nonlinear transport emerges from Kondo interaction, even without long-range magnetic order.

We also investigate correlation-driven nonlinear transport in $FeTe_{0.6}Se_{0.4}$ (see Supplementary Note 12). Similar to FeTe, nonlinear transport in $FeTe_{0.6}Se_{0.4}$ exhibits a twofold angular dependence, consistent with persistent short-range magnetic order [31]. Meanwhile, previous studies report an incoherent localized $d_{xy}$ orbital [46] and an orbital-selective Mott phase at elevated temperatures [27], which inhibits band hybridization at high temperatures. This is validated by nonlinear transport measurements as well: the third-order nonlinear transverse signals vanish above ~70 K. The agreement between correlation crossover and nonlinear transport in $FeTe_{0.6}Se_{0.4}$ supports our proposal of Kondo lattice-driven gap formation.

In addition to Kondo hybridization for gap opening and quantum metric effects, we consider other possible mechanisms as below. First, magnetic order can induce symmetry reduction and band folding, thus gap opening [47]. At low temperatures, the unit cell of FeTe expands due to lowered symmetry, causing Brillouin zone to shrink and bands to fold. Previous studies in Iron-Arsenide superconductors show that band folding via spin-density-wave can induce an energy gap [47]. However, in FeTe, the nonlinear transport persists above $T_N$, where neither lattice distortion nor crystal symmetry lowering occurs. Additionally, the third-order nonlinear transport is observed in $FeTe_{0.6}Se_{0.4}$ without static magnetic order. Second, as observed in topological insulators [48] and topological semimetals [49], spin-orbit coupling (SOC) modifies band structures through band splitting and gap opening. While SOC stabilizes the in-plane bicollinear antiferromagnetic order in FeTe and generates band splitting [29], it induces energy gap that is far away from Fermi level and in the size of ~50 meV [50] – resulting in negligible quantum metric effect. Moreover, SOC cannot explain the disappearance of nonlinear response at high temperatures, which can be naturally inferred from crossover in correlation effects.

Then we perform theoretical calculations of QMQ in FeTe and do a quantitative comparison with experiments. Based on Kondo hybridization scenario, we construct a tight-binding Hamiltonian up to a mean-field approximation describing interactions between Fe $3d_{xy}$ orbitals and Te $5p_z$ orbitals in one sublattice framework (Fig. 4a):

$$H_{\mathrm{Kondo}} = \sum_{ij} \Delta c_{ij}^{\dagger} f_{i-1,j} + \Delta' c_{ij}^{\dagger} f_{i+1,j} + \Delta c_{ij}^{\dagger} f_{i,j-1} + \Delta c_{ij}^{\dagger} f_{i,j+1} + \text{c.c.}, \tag{5}$$

where $\Delta$ and $\Delta'$ are antiferromagnetic-like and ferromagnetic-like interaction strengths, respectively, $c_{ij}$ and $f_{ij}$ are annihilation operators of itinerant Te orbitals and localized Fe orbitals, respectively. From a simplified 3-band model in the 2-Fe unit cell (see details in Methods section and Supplementary Note 13), we calculate (energy normalized) quantum metric tensor $G_{\mu\nu}$ (Figs. 4d-f) and the corresponding QMQ near the Fermi pockets around Γ and M region. Using semi-classical transport theory, we derive third-order nonlinear conductivity at the lower symmetry pocket around Γ and M point, reproducing experimental angular dependence (Fig. S31). The calculated QMQ contributions $\eta_{\parallel} \sim 2.11 \times 10^{-11}\ \mathrm{m^2V^{-2}}$ and $\eta_{\perp} \sim 4.02 \times 10^{-13}\ \mathrm{m^2V^{-2}}$ along longitudinal and transverse directions, respectively (see Supplementary Note 13), which has qualitative consistency with experimental values: $2.6 \times 10^{-11}\ \mathrm{m^2V^{-2}}$ and $-3.5 \times 10^{-12}\ m^2V^{-2}$ (Figs. 3c-d). Although Kondo correlation successfully describes the emergence of QMQ, there could be existing interplay between SOC and correlation effects like Hund's coupling, which may influence Fermi surface topology and magnetic order [51]' [52] and contribute nonlinear transport as well.

## Discussion and outlook

Our work demonstrates the probing of quantum geometry via nonlinear transport in a correlated system while preserving both *P* and *T*. As shown Fig. 4g, we summarize materials with different symmetries in previous literatures, whose quantum geometry is accessed by nonlinear transport. Over past years, extensive studies have been made in non-interacting systems and found that QMQ is not sensitive to *P* nor *T* whereas other high-order quantum geometry terms require symmetry breaking in *P* or *T* [3–6,12,16,17,19,21,53–61]. While progress has been made in sensing correlation effects and topological transitions by nonlinear transport, it is restricted in materials utilized strain or low symmetry to break *P* [17,18,24,25]. In this work, QMQ in FeTe reveals a strong electronic correlation can drive the quantum metric effect even with a high symmetry, which represents advances over previous studies in correlated systems. We note that QMQ persists in unstrained twisted graphene systems, consistent with this paradigm. Our work suggests a universal approach to detect correlation effects and phase transition through exploiting nonlinear transport and QMQ.

In conclusion, we demonstrate how correlation-driven quantum geometry induces nonlinear transport in antiferromagnetic FeTe. Our transport experiments reveal the third-order nonlinear transport at low temperatures, which originates from QMQ. Notably, the nonlinear transport of itinerant electrons reflects the symmetry reduction by localized magnetic moments, where the observed two-fold angular dependence in both monoclinic and tetragonal phases reflects bicollinear antiferromagnetic order and short-range magnetic fluctuations, respectively. We further show that a phenomenological Kondo-hybridization model can generate an energy gap and sizable QMQ. Based on an effective Kondo-lattice Hamiltonian, the calculated QMQ-induced nonlinear response is comparable to the value extracted from experiments. This demonstrates that strong quantum metric effects can emerge from strong electron correlations. In Supplementary Note 14, we compare our transport results and other characterization results from previous reports to show the interplay among nonlinear transport, Kondo physics and magnetic order. Under the case the high symmetry and vanishing net magnetic moment can hinder the reading of magnetic structure in collinear antiferromagnets [62], [63], our work proposes a zero-field electrical sensing method for Néel order and antiferromagnetic transitions via nonlinear transport. More crucially, these results provide a microscopic picture of nonlinear transport in a Kondo lattice system and advance quantum geometry engineering through tailored correlation effects.

## Figures and captions

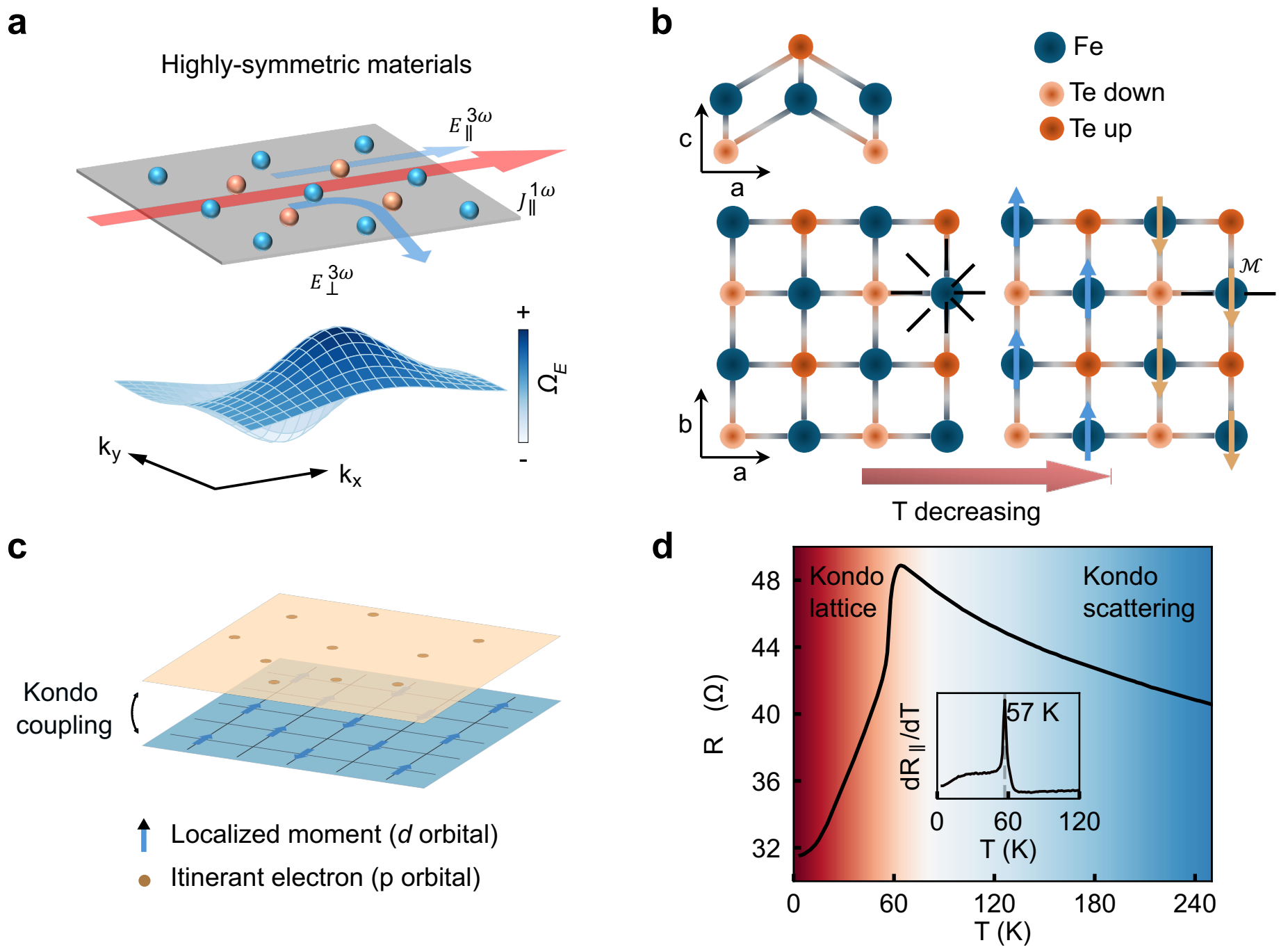

**Figure 1. Schematics of correlation-driven quantum geometry and FeTe. a,** Nonlinear transport as a probe for quantum geometry. In the upper panel, we show an example that QMQ can enable third-order nonlinear transport in materials with both $\mathcal{P}$ and $\mathcal{T}$ symmetries. Lower panel: the electric field-induced Berry curvature $\Omega_E$ in the momentum space under finite QMQ. **b,** Crystal structures of FeTe in tetragonal (lower left panel) and monoclinic (upper right panel) phases. The dashed line $\mathcal{M}$ denotes the mirror line. **c,** Schematic of Kondo interaction between d-orbital localized moments and p-orbital itinerant electrons. **d,** Measured longitudinal resistance and Kondo lattice-Kondo scattering regime transition in our FeTe sample. Inset shows the first derivative of resistance with respect to temperature as a function of temperature, where the peak in $dR_{\parallel}(T)/dT$ at 57 K is determined as the onset of antiferromagnetic order at $T_N$.

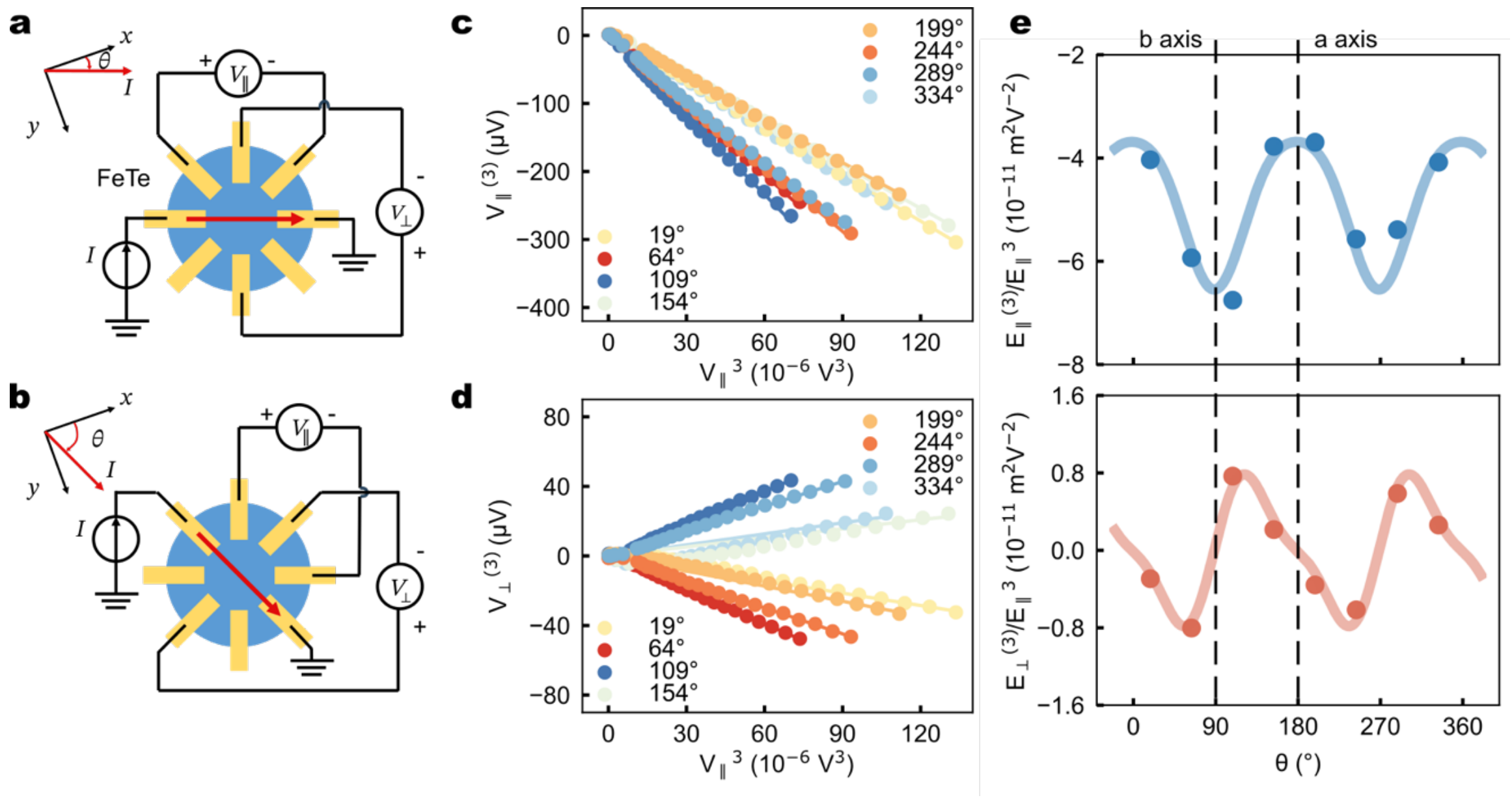

**Figure 2. Observation of the third-order nonlinear transport in FeTe. a, b,** Electrode configuration for recording longitudinal and transverse voltages. By rotating the contact configuration, angle-dependent nonlinear response is measured. **c, d,** $V_{\parallel}^{(3)}$**(c)** and $V_{\perp}^{(3)}$**(d)** as a function of $V_{\parallel}^{3}$ when current is applied along different directions ($\theta$) at 20 K (below $T_N$). The symbols are experimental data, and the lines are linear fits. **e,** $E_{\parallel}^{(3)}/E_{\parallel}^{3}$ and $E_{\perp}^{(3)}/E_{\parallel}^{3}$ as a function of $\theta$ at 20 K. The symbols represent experimentally determined values and solid lines are fitted curves using Eqs. (1) and (2).

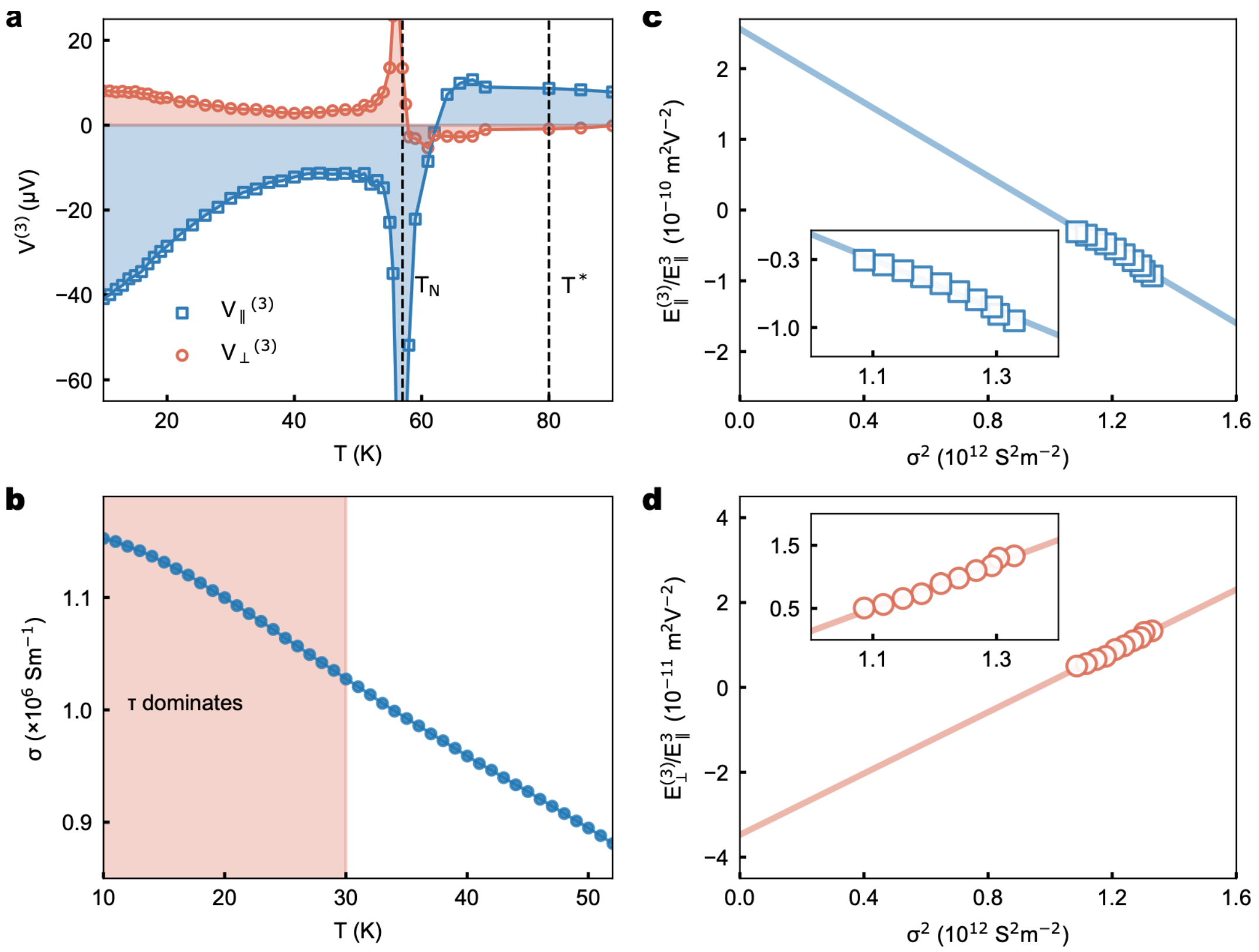

**Figure 3. Nonlinear transport from quantum metric quadrupole**. **a,** the third-order longitudinal ($V_{\parallel}^{(3)}$) and transverse ($V_{\perp}^{(3)}$) signals measured at different temperatures. **b,** Conductivity ($\sigma$) as a function of temperature. **c, d,** Scaling law analysis. $E_{\parallel}^{(3)}/E_{\parallel}^{3}$ **(c)** and $E_{\perp}^{(3)}/E_{\parallel}^{3}$ **(d)** as a function of $\sigma^2$. Insets show good consistency between our experimental data and linear fitting.

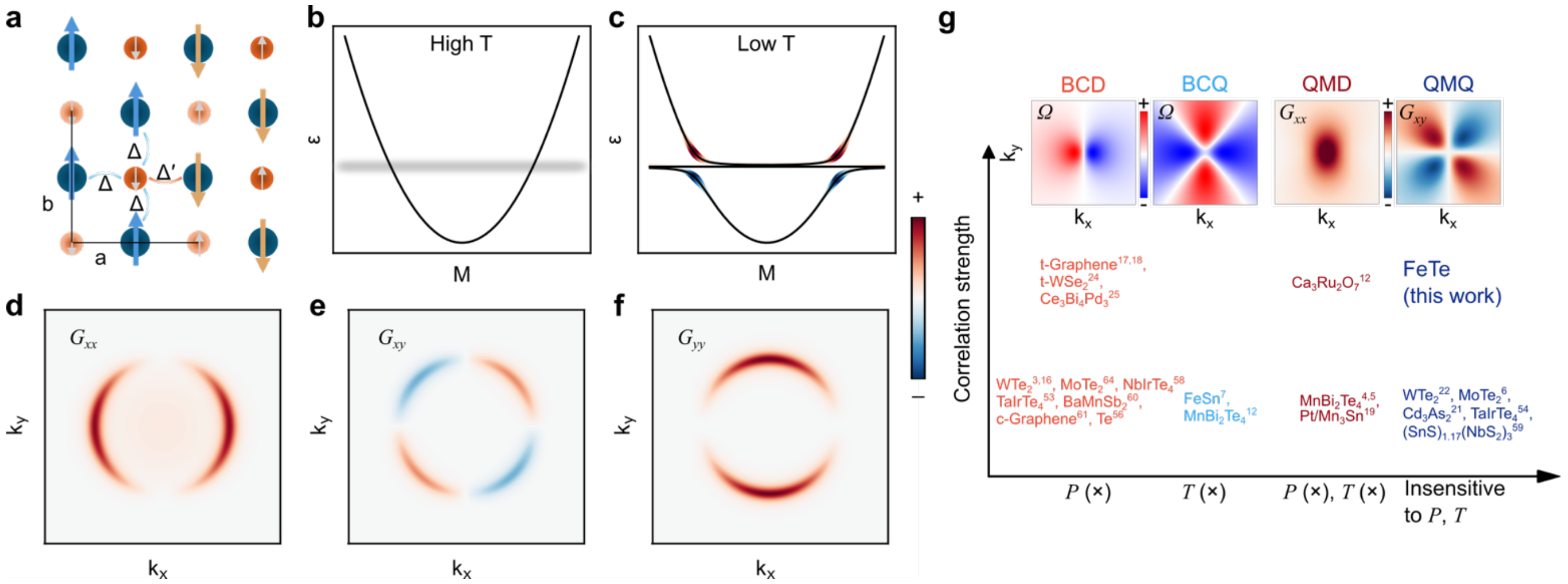

**Figure 4. Energy hybridization and quantum metric distribution under Kondo correlation. a,** Schematic of Kondo interaction: Fe atoms (dark blue balls with arrows) interact with Te atoms (red balls with arrows) via hybridization between Fe $3d_{xy}$ and Te $5p_z$ orbitals (arcs). **b, c,** Schematic band structures near the M point in the phenomenological Kondo-hybridization model at high **(b)** and at low **(c)** temperatures. In **b**, the grey band denotes localized Fe $3d_{xy}$ orbital and the black band denotes Te $5p_z$ orbital. In **c**, distribution of normalized quantum metric $G_{xx}$ is shown. **d-f,** Calculated normalized quantum metric $G_{\mu\nu}$ distribution for upper hybridized bands in **c** ($G_{xx}$ **(d)**, $G_{xy}$ **(e)** and $G_{yy}$ **(f)**) in the $k_z = 0$ plane of momentum space. **g,** Summary of quantum geometry sensed by nonlinear transport in different materials [3,4,22,24,25,53,54,56,58,60,61,64,5–7,16–19,21]. c-Graphene refers to corrugated graphene. Berry curvature is calculated from tilted massive Dirac models under different symmetry constraints and plotted to show BCD and BCQ pattern in momentum space, respectively. Normalized quantum metric component $G_{xx}$ and $G_{xy}$ are calculated from massive Dirac model under different symmetry constraints and plotted to show QMD and QMQ patterns in momentum space, respectively.

## Methods

**Sample fabrication.** High-quality bulk FeTe crystals were purchased from 2D semiconductor Inc. Our FeTe device fabrication is described as follows. First, FeTe flakes were exfoliated from bulk material to Si/$SiO_2$ (285 nm thickness) wafer inside a glovebox with a $N_2$ environment ($H_2O$ and $O_2$ levels < 0.01 ppm). Then e-beam lithography and metal deposition (Ti (10nm)/Au (100 nm)) were performed using thermal evaporation. Finally, the device was encapsulated with hBN to prevent FeTe degradation during measurements.

**Electrical measurement.** The transport measurements were performed in Cryogenic Free Measurement System (Cryogenic Ltd.). A sinusoidal ac current in frequency of 19.357 Hz was applied using Keithley 6221 current source. Longitudinal and transverse voltages were recorded using Stanford SR830 lock-in amplifiers. Signal purity was verified in Supplementary Note 2.

**Band structure calculations.** DFT calculations were performed using Vienna ab initio simulation package [65]. The projector-augmented wave potential was adopted with the plane-wave energy cutoff set at 500 eV, and the convergence criteria as $10^{-5}$ eV. The exchange-correlation functional of the generalized gradient approximation (GGA) has been used, with a 6 × 8 × 6 gamma-centered Monkhorst–Pack mesh. The GGA+U method is employed to treat the strong correlations of the Fe 3d orbitals. Band-unfolding was applied to obtain the effective band structure in the Brillouin zone without magnetic order [66]' [67].

**The theoretical estimate of third-order nonlinear response based on Kondo lattice.** The Kondo lattice model used in this study describes the interaction between Fe $3d_{xy}$ orbital and its neighboring Te $5p_z$ orbital. The tight-binding model adopts a 2-Fe unit cell, consisting of two Fe $3d_{xy}$ orbitals and two Te $5p_z$ orbitals, forming two sublattices with up and down Te atoms. The interaction Hamiltonians of the two sublattices are written as,

$$H_{\mathrm{Kondo}}^{\mathrm{u}} = \sum_{ij} \Delta c_{ij}^{\dagger} f_{i-1,j} + \Delta' c_{ij}^{\dagger} f_{i+1,j} + \Delta c_{ij}^{\dagger} f_{i,j-1} + \Delta c_{ij}^{\dagger} f_{i,j+1} + \text{c.c.},$$

$$H_{\mathrm{Kondo}}^{\mathrm{d}} = \sum_{ij} \Delta c_{ij}^{\dagger} f_{i-1,j} + \Delta c_{ij}^{\dagger} f_{i+1,j} + \Delta c_{ij}^{\dagger} f_{i,j-1} + \Delta' c_{ij}^{\dagger} f_{i,j+1} + \text{c.c.},$$

In the k-space, the interaction Hamiltonian is

$$H_{\mathrm{int}}(\mathbf{k}) = \sum_{\mathbf{k}} 2\Delta \cos\frac{1}{2} bk_y \; c_{1\mathbf{k}}^{\dagger} f_{1\mathbf{k}} + \left[(\Delta + \Delta')\cos\frac{1}{2} ak_x - i(-\Delta + \Delta')\sin\frac{1}{2} ak_x\right] c_{1\mathbf{k}}^{\dagger} f_{1\mathbf{k}} + \text{c.c.}$$

$$+ 2\Delta \cos\frac{1}{2} bk_y \; c_{2\mathbf{k}}^{\dagger} f_{2\mathbf{k}} + \left[(\Delta + \Delta')\cos\frac{1}{2} ak_x + i(-\Delta + \Delta')\sin\frac{1}{2} ak_x\right] c_{2\mathbf{k}}^{\dagger} f_{2\mathbf{k}} + \text{c.c.}$$

with subscripts 1 and 2 being the two sublattices. We can write it into a matrix form,

$$H_{\mathrm{int}}(\mathbf{k}) = \begin{pmatrix} 0 & 0 & w & x - iy \\ 0 & 0 & x + iy & w \\ w & x - iy & 0 & 0 \\ x + iy & w & 0 & 0 \end{pmatrix}$$

under basis $\psi = (c_1\ c_2\ f_1\ f_2)^T$ with $w = 2\Delta\cos\frac{1}{2}bk_y$, $x = (\Delta + \Delta')\cos\frac{1}{2}ak_x$ and $y = (-\Delta + \Delta')\sin\frac{1}{2}ak_x$.

As Fe $3d_{xy}$ orbital is localized, we assume its energy dispersion is a trivial nearly flat band $\epsilon_d = 0$. On the contrary, the Te $5p_z$ orbitals of the two sublattices should form two bands $\epsilon_+(\mathbf{k})$ and $\epsilon_-(\mathbf{k})$, with eigenstates being $\frac{1}{\sqrt{2}}(c_1 + c_2)$ and $\frac{1}{\sqrt{2}}(c_1 - c_2)$. The Hamiltonian with diagonalized Te $5p_z$ orbitals is,

$$\widetilde{H}(\mathbf{k}) = \begin{pmatrix} \epsilon_+ & 0 & w+x & iy \\ 0 & \epsilon_- & -iy & w-x \\ w+x & iy & 0 & 0 \\ -iy & w-x & 0 & 0 \end{pmatrix}$$

We further assume that $\epsilon_+$ has much higher energy and thus can be ignored. Then we only focus on the reduced 3-band Hamiltonian,

$$\widetilde{H}_{\mathrm{r}}(\mathbf{k}) = \begin{pmatrix} \epsilon_- & -iy & w-x \\ iy & 0 & 0 \\ w-x & 0 & 0 \end{pmatrix}.$$

The eigenvalues and eigenvectors of this matrix is,

$$\lambda = 0,\ \frac{1}{2}\left(\epsilon_- \pm \sqrt{\epsilon_-^2 + 4m^2}\right), \quad m^2 = y^2 + (w-x)^2,$$

$$\lambda = 0,\ \psi_0 = \alpha_0 \begin{pmatrix} 0 \\ -x \\ iy \end{pmatrix},$$

$$\lambda = \lambda_\pm,\ \psi_\pm = \alpha_\pm \begin{pmatrix} \lambda_\pm \\ iy \\ w-x \end{pmatrix}.$$

The third order conductivity from quantum metric is computed by the equation,

$$\sigma_{\mu\alpha\beta\gamma}^{\mathrm{QM}(3)}(3\omega) = \frac{e^4}{4\hbar^2}\tau \int \frac{\mathrm{d}\boldsymbol{k}^2}{(2\pi)^2}\left[\left(\partial_\alpha\partial_\beta G_{\mu\gamma} - \partial_\mu\partial_\beta G_{\alpha\gamma} + \partial_\mu\partial_\alpha G_{\beta\gamma}\right)f_0 - \frac{\hbar^2}{2} v_\mu v_\alpha G_{\beta\gamma}\frac{\partial^2 f_0}{\partial\epsilon^2}\right].$$

Where normalized quantum metric $G_{\mu\nu}^n = 2\mathrm{Re}\sum_{m\neq n}\frac{\mathcal{A}_\mu^{nm}\mathcal{A}_\nu^{mn}}{\epsilon_{n\mathbf{k}}-\epsilon_{m\mathbf{k}}}$. $\mathcal{A}_\mu^{nm}$ is the inter-band Berry connection defined as $\mathcal{A}_\mu^{nm} = \left\langle u_{\boldsymbol{k}}^n \middle| i\frac{\partial}{\partial k_\mu} \middle| u_{\boldsymbol{k}}^m\right\rangle$. Numerical integration is then computed on this equation to obtain nonlinear response tensor.

## Data availability

The data that support the plots within this paper and other findings of this study are available at XXX. (Authors' note: the data will be uploaded after the acceptance of this manuscript.)

## Code availability

The simulation code used for this study can be found at XXX. (Authors' note: the codes will be uploaded after the acceptance of this manuscript.)

## Acknowledgements

We thank Xue He and Rui Wu for technical assistance with FeTe etching and helpful discussions. The authors acknowledge funding support from the National Key R&D Program of China (grant number 2021YFA1401500), the NSFC/RGC Joint Research Scheme (N_HKUST620/21 and 52161160334), and the Research Grant Council (grant number 16303322, 16303521, 16309924 and T-46-705/23-R). This research was partially supported by the State Key Laboratory of Advanced Displays and Optoelectronics Technologies.

## Author Contributions

Q.S. and R.L. conceived the experiments. R.L., X.W., K.Q., and C.H.C. fabricated the device. R.L. conducted electrical transport measurements. Z.C. carried out the tight-binding model calculations. X.C. carried out density functional theory analysis. X.R. performed atomic force microscopy measurements. Y.Z. performed transmission electron microscopy characterization. R.L., C.Z. and X.C. analyzed the data. R.L. and Z.C. drafted the manuscript and all the authors reviewed the manuscript.

## Competing Interest Declaration

The authors declare no competing financial interest.